\documentstyle[12pt,epsf]{article}
\newcommand{\be}{\begin{equation}}
\newcommand{\ee}{\end{equation}}
\newcommand{\bea}{\begin{eqnarray}}
\newcommand{\eea}{\end{eqnarray}}
\newcommand{\nn}{\nonumber}

\begin{document}

\def\gamh{\Gamma_H}
\def\eb{E_{\rm beam}}
\def\deb{\Delta E_{\rm beam}}
\def\sigm{\sigma_M}
\def\sigmmax{\sigma_M^{\rm max}}
\def\sigmmin{\sigma_M^{\rm min}}
\def\sige{\sigma_E}
\def\dsigm{\Delta\sigma_M}
\def\mh{M_H}
\def\lyear{L_{\rm year}}

\def\wstar{W^\star}
\def\zstar{Z^\star}
\def\ie{{\it i.e.}}
\def\etal{{\it et al.}}
\def\eg{{\it e.g.}}
\def\pzero{P^0}
\def\mt{m_t}
\def\mpzero{M_{\pzero}}
\def\mev{~{\rm MeV}}
\def\gev{~{\rm GeV}}
\def\gam{\gamma}
\def\lsim{\mathrel{\raise.3ex\hbox{$<$\kern-.75em\lower1ex\hbox{$\sim$}}}}
\def\gsim{\mathrel{\raise.3ex\hbox{$>$\kern-.75em\lower1ex\hbox{$\sim$}}}}
\def\ntc{N_{TC}}
\def\epem{e^+e^-}
\def\tauptaum{\tau^+\tau^-}
\def\lplm{\ell^+\ell^-}
\def\anti{\overline}
\def\mw{M_W}
\def\mz{M_Z}
\def\fbi{~{\rm fb}^{-1}}
\def\mupmum{\mu^+\mu^-}
\def\rts{\sqrt s}
\def\sigrts{\sigma_{\tiny\rts}^{}}
\def\sigrtssq{\sigma_{\tiny\rts}^2}
\def\sigrtsprime{\sigma_{E}}
\def\nsigrts{n_{\sigrts}}
\def\gampzero{\Gamma_{\pzero}}
\def\pzerop{P^{0\,\prime}}
\def\mpzerop{M_{\pzerop}}

\font\fortssbx=cmssbx10 scaled \magstep2
\hbox to \hsize{
%
%\special{psfile=uwlogo.ps
% hscale=8000 vscale=8000
% hoffset=-12 voffset=-2}
%\hskip.5in \raise.1in
%
$\vcenter{
\hbox{\fortssbx University of Florence}
\hbox{\fortssbx University of Geneva}
}$
\hfill
$\vcenter{
\hbox{\bf DFF-339/5/99}
\hbox{\bf UGVA-DPT-1999 05-1044}
}$
}
%}

%
\medskip
\begin{center}

{\Large\bf\boldmath Bounds on New Physics from the New Data on Parity
Violation in Atomic Cesium\\}
\rm
\vskip1pc
{\Large R. Casalbuoni$^{a,b}$,  S. De Curtis$^b$,
\\ D. Dominici$^{a,b}$ and R. Gatto$^c$\\}
\vspace{5mm}
{\it{$^a$Dipartimento di Fisica, Universit\`a di Firenze, I-50125
Firenze, Italia
\\
$^b$I.N.F.N., Sezione di Firenze, I-50125 Firenze, Italia\\
$^c$D\'epart. de Physique Th\'eorique, Universit\'e de
Gen\`eve, CH-1211 Gen\`eve 4, Suisse}}
\end{center}
\bigskip
\begin{abstract}
\noindent We assume the latest experimental determination of
the weak charge of atomic cesium and analyze its implications for possible
new physics. We notice that the data would imply  positive upper and lower
bounds on the new physics contribution to the weak charge, $\delta_NQ_W$.
The required new physics
should be of a type not severely constrained by the high energy precision data.
A simplest possibility would be new neutral vector bosons almost
 un-mixed to the $Z$ and with sizeable couplings to fermions. The
lower positive bound would however forbid zero or negative
$\delta_NQ_W$ and exclude at 99\% CL not only the standard model
but also models with sequential $Z^\prime$, in particular
simple-minded towers of $Z$-like excitations from
extra-dimensions. The bound would also imply an upper limit on
the $Z^\prime$ mass within the models allowed. Conclusions are
also derived for models of four-fermion contact interactions.
\end{abstract}
\newpage
In a recent paper \cite{bennett} a new determination of the weak
charge of atomic cesium has been reported. It takes
advantage from a new measurement of the tensor transition
probability for the $6S\to 7S$ transition in cesium and from
improvement of the atomic structure calculations in light of new
experimental tests. The value reported
\be
Q_W(^{133}_{55}Cs)=-72.06\pm (0.28)_{\rm expt}\pm (0.34)_{\rm
theor}
\label{newexp}
\ee
represents a considerable improvement with respect to the previous
determination \cite{noecker,blundell}
\be
Q_W(^{133}_{55}Cs)=-71.04\pm (1.58)_{\rm expt}\pm (0.88)_{\rm
theor}
\ee
In the following we assume the validity of the determination in eq.
(\ref{newexp}),
{\it verbatim et literatim}.
 We note that the new measure of $Q_W$ is at a level ($\approx
.6 \%$) comparable to the precisions attained for electroweak observables
in high energy experiments.

On the theoretical side, $Q_W$ can be expressed as
\cite{marciano}
\be
Q_W=-72.84\pm 0.13-102\epsilon_3^{\rm rad}+\delta_NQ_W
\ee
including hadronic-loop uncertainty. We use here the variables
$\epsilon_i$ (i=1,2,3) of ref. \cite{altarelli}, which include the
radiative corrections, in place of the set of
variables $S$, $T$ and $U$ originally introduced in ref. \cite{peskin}.
In the above
definition of $Q_W$
we have  explicitly included only the Standard Model (SM) contribution
to the radiative corrections. New physics (that is physics beyond the
SM) contributions
to $\epsilon_3$ are represented by the term $\delta_N Q_W$. Also, we
have neglected a correction proportional to $\epsilon_1^{\rm rad}$.
In fact, as well known \cite{marciano}, due to the
particular values of the number of neutrons ($N=78$) and of
protons ($Z=55$) in cesium, the dependence on
$\epsilon_1$ almost cancels out.

From the theoretical expression
we see that $Q_W$ is particularly sensitive to new physics
contributing to the parameter $\epsilon_3$. This kind of new
physics is severely constrained by the high energy experiments.
From a recent analysis \cite{altarelli2}, one has that
the  value of $\epsilon_3$ from the  high
energy data is
\be
\epsilon_3^{\rm expt}=(4.1\pm 1.4)\times 10^{-3}
\ee
To estimate new physics contributions to this parameter one
has to subtract the SM radiative corrections, which, for
$m_{top}=175~GeV$ and for $m_H=70,~300,~1000~GeV$, are given
respectively by (we use a linear interpolation of the results of
ref. \cite{altarelli2})
\be
\matrix{m_H=70~GeV && \epsilon_3^{\rm rad}=4.925\times 10^{-3}\cr
        m_H=300~GeV && \epsilon_3^{\rm rad}=6.115\times
        10^{-3}\cr m_H=1000~GeV && \epsilon_3^{\rm
        rad}=6.65\times 10^{-3}}
\ee
Therefore new physics contributing to $\epsilon_3$
cannot be larger than a few per mill. Since $\epsilon_3$ appears
in $Q_W$ multiplied by a factor 102, this kind of new physics which
contributes through $\epsilon_3$
cannot contribute to $Q_W$ for more than a few tenth. On the
other side the discrepancy between the SM and the experimental
data is given by (for a light Higgs)
\be
Q_W^{\rm expt}-Q_W^{SM}=1.28\pm 0.46
\ee
where we have added in quadrature the uncertainties. In order to
reproduce such an effect with new physics contributing to
$\epsilon_3$ we would need $\epsilon_3=(-12.5\pm 4.5)\times
10^{-3}$, just an order of magnitude bigger than what possibly allowed
from high energy. Therefore, if we believe that the latest experiment and
the theoretical evaluations of the atomic structure effects are
correct,   we need  new physics of a type which is not
constrained by the high energy data.

Before discussing this item
let us see which bounds the data put on $\delta_N Q_W$.
The deviation from the data is a function of $m_H$ given by
\be
\Delta Q_W=Q_W^{\rm expt}-Q_W(m_H)=
0.78+102\epsilon_3^{\rm rad}-\delta_N Q_W
\ee
with an uncertainty of $0.46$. We thus get the following
bounds
\be
(0.78+102\epsilon_3^{\rm rad})-0.46 c\le \delta_N Q_W\le
(0.78+102\epsilon_3^{\rm rad})+0.46 c
\ee
where $c$ specifies the confidence level, ($c=1.96$ for 95\% CL and
$c=2.58$ for 99\% CL). Notice that $\epsilon_3^{\rm rad}$ increases
with $m_H$ and therefore the upper and lower bounds also
increase with $m_H$. For example, for the choice $m_H=70~GeV$ (the direct
experimental bound on $m_H$ is larger than this value, but this
is just to make an example of what is going on), we get
\bea
95\%~ {\rm CL}, && 0.38\le\delta_NQ_W\le 2.18\nn\\ 99\%~ {\rm
CL}, && 0.10\le\delta_N Q_W\le 2.47
\label{bounds}
\eea
whereas for $m_H=300~GeV$ the bounds are
\bea
95\%~ {\rm CL}, && 0.50\le\delta_NQ_W\le 2.30\nn\\ 99\%~ {\rm
CL}, && 0.22\le\delta_N Q_W\le 2.59
\label{bounds2}
\eea
In particular we see that a  negative or zero contribution to
$Q_W$ from new physics is excluded at 99\% CL. In view of the
preceding observation this result gets  stronger when increasing
the Higgs mass. Therefore, if one assumes that the new result on
atomic cesium is not due to some statistical fluctuation and that
the theoretical errors have not been underestimated in some way,
one would formally be lead to conclude that the SM is excluded at
99\% CL.

Let us now look at models which, at least in principle, could
give rise to a sizeable modification of $Q_W$. In ref.
\cite{altarelli3} it was pointed out that models involving extra
neutral vector bosons coupled to ordinary fermions can do the
job ( there is an extensive literature on the phenomenology of additional
Z- bosons \cite{Z} ). In fact it was shown that the corrections to $Q_W$ are
given in these models by
\bea
\delta_NQ_W&=&
16\Big\{\frac 1
{16}\left[\left(1+4\frac{s_{\theta}^4}{c_{2\theta}}\right)Z-N\right]
\Delta\rho_M\nn\\
&+&\left[\left(2Z+N\right)\left(a_e v_u^\prime+a_e^\prime
v_u\right)+\left(Z+2N\right)\left(a_e v_d^\prime+a_e^\prime
v_d\right)\right]\xi\nn\\ &+&\left[\left(2Z+N\right)a_e^\prime
v_u^\prime+
\left(Z+2N\right) a_e^\prime v_d^\prime\right]
\frac{M_Z^2}{M_{Z^\prime}^2}\Big\}
\label{deltaQ}
\eea
where $\xi$ is the mixing angle,  $a_f, v_f, a_f^\prime,
v_f^\prime$ are the couplings of $Z$ and $Z^\prime$ to fermions,
and $\Delta\rho_M$ is an additional contribution to the $\rho$
parameter arising from the mixing \cite{altarelli3}
\be
\Delta\rho_M=\sin^2\xi\left(\frac{M_{Z^\prime}^2}{M_{Z}^2}-1\right)
\ee
The first term in $\delta_N Q_W$ arises from the modification of
the $\rho$ parameter, which induces also a redefinition of
$s_\theta^2$. The second is due to the $Z-Z^\prime$ mixing,
whereas the third one is due to the direct exchange of the
$Z^\prime$.

The high energy data at the Z resonance strongly bound the first two
terms which depend on the mixing, but they say nothing about the
third contribution. Therefore, in order to get an insight about
the order of magnitude of the effects due to this kind of new
physics which is not mainly expressed in the mixing, we will concentrate
only on the last term, since we
already know that the first two terms would be unable to give
corrections to $Q_W$ of the right order of magnitude. For this
reason we will assume for simplicity in the following calculations $\xi=0$.
In this
case $\delta_NQ_W$ is completely fixed by the $Z^\prime$
parameters: its couplings to the electron and to the up and down
quark, and its mass.

We will discuss three classes of models: the left-right (LR) models,
the extra-U(1) models,  and the so-called sequential SM
models  (that is models with fermionic couplings just scaled from those
 of the SM). The relevant couplings are given in Table 1.

\begin{table}[bht]
\caption{Vector and axial-vector coupling constants for the determination of
$\delta_NQ_W$ for the various models considered in the text. The
SM couplings are for the sequential SM. The different
extra-U(1) models are parameterized by the angle $\theta_2$, and
in the table $c_2=\cos\theta_2$, $s_2=\sin\theta_2$. This angle
takes a value between $-\pi/2$ and $+\pi/2$.}
\begin{center}
\begin{tabular}{|c|c|c|}
\hline
SM & Extra-U(1) & LR \\
\hline\hline
$a_e=\frac 1 4 $& $a_e^\prime=\frac 1 4 s_\theta\left(-\frac 1 3
c_2+\sqrt{\frac 5 3} s_2\right)$&$a_e^\prime=-\frac 1 4
\sqrt{c_{2\theta}}$\\
\hline
$v_u=\frac 1 4 - \frac 2 3 s_\theta^2 $&$ v_u^\prime=0 $&$
v_u^\prime=\left(\frac 1 4 - \frac 2 3
s_\theta^2\right)/\sqrt{c_{2\theta}}$\\
\hline
$v_d =-\frac 1 4 + \frac 1 3 s_\theta^2$ &
 $v_d^\prime=\frac 1 4 s_\theta\left( c_2+\sqrt{\frac 5 3}
s_2\right)$ &$  v_d^\prime=\left(-\frac 1 4 +
\frac 1 3 s_\theta^2\right)/\sqrt{c_{2\theta}}$\\
\hline
\end{tabular}
\end{center}
\end{table}

In the case of the  LR model \cite{LR} we get a contribution
\be
\delta_NQ_W=-\frac{M_Z^2}{M_{Z^\prime}^2}Q_W^{SM}~~~~
\ee
For this model one has a 95\% lower bound on $M_{Z^\prime}$ from
Tevatron \cite{tevatron} given by $M_{Z^\prime}\ge 630~GeV$
implying, for $70\le m_H(GeV)\le 1000$, $\delta_NQ_W\le 1.54$.
Therefore a LR model could explain the data allowing for a mass
of the $Z^\prime$ varying between the intersection from the 95\%
CL bounds deriving from eq. (\ref{bounds})
\bea
m_H=70~GeV && 529\le M_{Z^\prime}(GeV)\le 1267\nn\\ m_H=300~GeV
&& 515\le M_{Z^\prime}(GeV)\le 1105\nn\\
\eea
and the lower bound of $630~GeV$ .

In the case of the extra-U(1) models \cite{U(1)} the experimental
bounds for the masses vary according to the values of the
parameter $\theta_2$ (see Table 1), but in general they are about
$600~ GeV$ at 95\% CL \cite{tevatron,langa}. From eq.
(\ref{deltaQ}) with $\xi=0$ we can easily see that  the models
with $\theta_2$ in the interval   $-0.66\le\theta_2({\rm rad})\le
0.25$  give $\delta_NQ_W\le 0$, and therefore they are excluded
at the 99\% CL. In particular the model known in the literature
as the $\eta$ or $A$ model, which corresponds to $\theta_2=0$, is
excluded.

The bounds in eqs. (\ref{bounds})  and (\ref{bounds2}) at 95\% CL
can be translated into  lower and upper bounds on $M_{Z^\prime}$.
The result is given in Fig. 1, where the bounds are plotted
versus $\theta_2$. In looking at this figure one should also
remember that the direct lower bound from Tevatron is about 600
~GeV at 95\% CL, leading to an exclusion region
$-0.84\le\theta_2({\rm rad})\le 0.43$ for $m_H=70~GeV$ and
$-0.89\le\theta_2({\rm rad})\le 0.48$ for $m_H=300~GeV$.

The last possibility we consider is a sequential SM. In this case
we assume that the couplings are the ones of the SM just scaled
by a common factor $a$. Therefore we get
\be
\delta_NQ_W=a^2 \frac{M_Z^2}{M_{Z^\prime}^2}Q_W^{SM}
\label{qwkk}
\ee
We see that no matter what the choice of $a$ is, the sign of the new physics
contribution turns out to be negative. Therefore all this class
of models are excluded at 99\% CL.

This result can be trivially extended to certain models based on
extra dimensions which have a tower of Kaluza-Klein resonances of
the $Z$ all coupled as the $Z$, except for a factor $a=\sqrt{2}$
\cite{KK}. Taking also into account the modification of the Fermi
constant $G_F$ due to the KK excitations of the $W$, the eq.
(\ref{qwkk}) gets an additional positive factor $s_{\theta}^2$.
The effect of having an infinite tower, if any, worsens the
situation. For instance, in the case of  one extra-dimension, one
gets a further factor of $\pi^2/6$. Therefore the experimental
result on atomic parity violation in atomic cesium excludes at a
very high confidence level these simple-minded extra-dimension
models of additional bosons $Z^\prime$.

Another interesting possibility one can analyze is that of a four-fermion
contact interaction, which could arise from different theoretical origins.
Also this case has no visible effects at the
$Z$ peak. We will follow  the analysis and the notations of
ref. \cite{langacker}. In this situation it turns out to be
convenient to express the weak charge as
\be
Q_W=-2\left[c_{1u}(2Z+N)+c_{1d}(Z+2N)\right]
\ee
where $c_{1u,d}$ are products of vector and axial-vector
couplings. We will consider models with a contact
interaction given by
\be
{\cal L}=\pm \frac{4\pi}{\Lambda^2}\bar e\Gamma_\mu e\bar q\Gamma^\mu q,~~~~
\Gamma_\mu=\frac 1 2\gamma_\mu(1-\gamma_5)
\ee
This leads to a shift in the couplings given by
\be
c_{1u,d}\to c_{1u,d}+\Delta C,~~~~\Delta C=\mp\frac{\sqrt{2}\pi}{G_F\Lambda^2}
\ee
Since a variation of the couplings induces a variation of $Q_W$
of opposite sign, we see that the choice of the negative sign in
the contact interaction is excluded. In the case of the positive
sign, using  the 95\% CL bounds given in eq. (\ref{bounds})  we
get
\be
11.8\le\Lambda(TeV)\le 28.2
\ee

Let us now consider a contact interaction induced by
lepto-quarks. Following again ref. \cite{langacker}, we take the
case of so-called $SU(5)$-inspired leptoquarks, leading to the interaction
\be
{\cal L}=\frac{\eta_L^2}{2M_S^2}\bar e_L\gamma_\mu e_L\bar
u_L\gamma^\mu u_L+\frac{\eta_R^2}{2M_S^2}\bar e_R\gamma_\mu
e_R\bar u_R\gamma^\mu u_R
\ee
From the constraints on $\pi_{e2}/\pi_{\mu 2}$ one expects
$\eta_L\approx 0$ or $\eta_R\approx 0$. Only the coupling
$c_{1u}$ has a shift
\be
c_{1u}\to c_{1u}+\Delta C,~~~~
\Delta C= \mp\frac{\sqrt{2}\eta_{L,R}^2}{8G_F M_S^2}
\ee
It follows that the shift on $Q_W$ is negative for
$\eta_R\not=0$. Therefore only the left coupling is allowed
($\eta_R=0$). In that case we get the bounds (again from eq.
(\ref{bounds}))
\be
1.6\le \frac{M_S(TeV)}{\eta_L}\le 3.9
\ee
If one assumes $\eta_L^2\approx 4\pi\alpha$, it follows
\be
0.48\le M_S(TeV)\le 1.17
\ee
Constraints on four-Fermi contact interactions from low energy electroweak
experiments have been also considered in \cite{cho}.

In conclusion, we have assumed the validity of the  new determination
of the weak
charge of atomic cesium, which  represents a big improvement
with  respect to the previous result on this important quantity, and we have
analyzed some of its theoretical implications.  We have shown that the present
data imply  positive upper and lower bounds on the new physics
contribution to the weak charge, $\delta_NQ_W$. It turns out from our analysis
that  new
physics contributing to both  high and low energy precision electroweak data
cannot easily reproduce the experimental situation, due to the existing
experimental constraints from the high energy precision data. In principle
it is
possible to explain the new data with new neutral vector bosons
which are un-mixed (or almost so) with the $Z$ and have a
sizeable couplings to the fermions. In this way the $Z^\prime$
contributes only to the low energy physics, allowing, in
principle, to explain the new data on parity violation, without
contradicting the $Z$-physics results from LEP and SLD. The lower
positive bound on such new physics effects to the weak charge would have
striking consequences. In fact it would forbid zero or negative
$\delta_NQ_W$ at more than 99\% CL. That is, if one takes
 the data seriously, the SM and  extra-vector boson
models with $\delta_NQ_W<0$ would seem to be excluded at 99\% CL.
In particular,
models with $Z^\prime$ couplings to fermions obtained via a
simple scaling from the SM couplings (sequential SM) would be
excluded. In fact, since $Q_W<0$,  they would give a
negative contribution. Therefore, also simple-minded models from
extra-dimensions describing a tower of $Z$-like excitations would be
excluded at 99\% CL.
 Also, the existence of the  positive lower bound would imply
for the allowed models an upper limit on the $Z^\prime$ mass. We have
also analyzed the implications of the new data for possible four-fermion
contact interactions of different origins.

Of course, one should avoid drawing general conclusions, since more data
will be necessary and because only some typical models have been
considered here. We can only repeat, with Gustave Flaubert, in his
correspondence with Louis Bouilhet: {\it Nous sommes un fil et nous
voulons savoir la trame. Contentons-nous du tableau, c'est aussi bon}.

\begin{figure}
\centerline{\epsffile{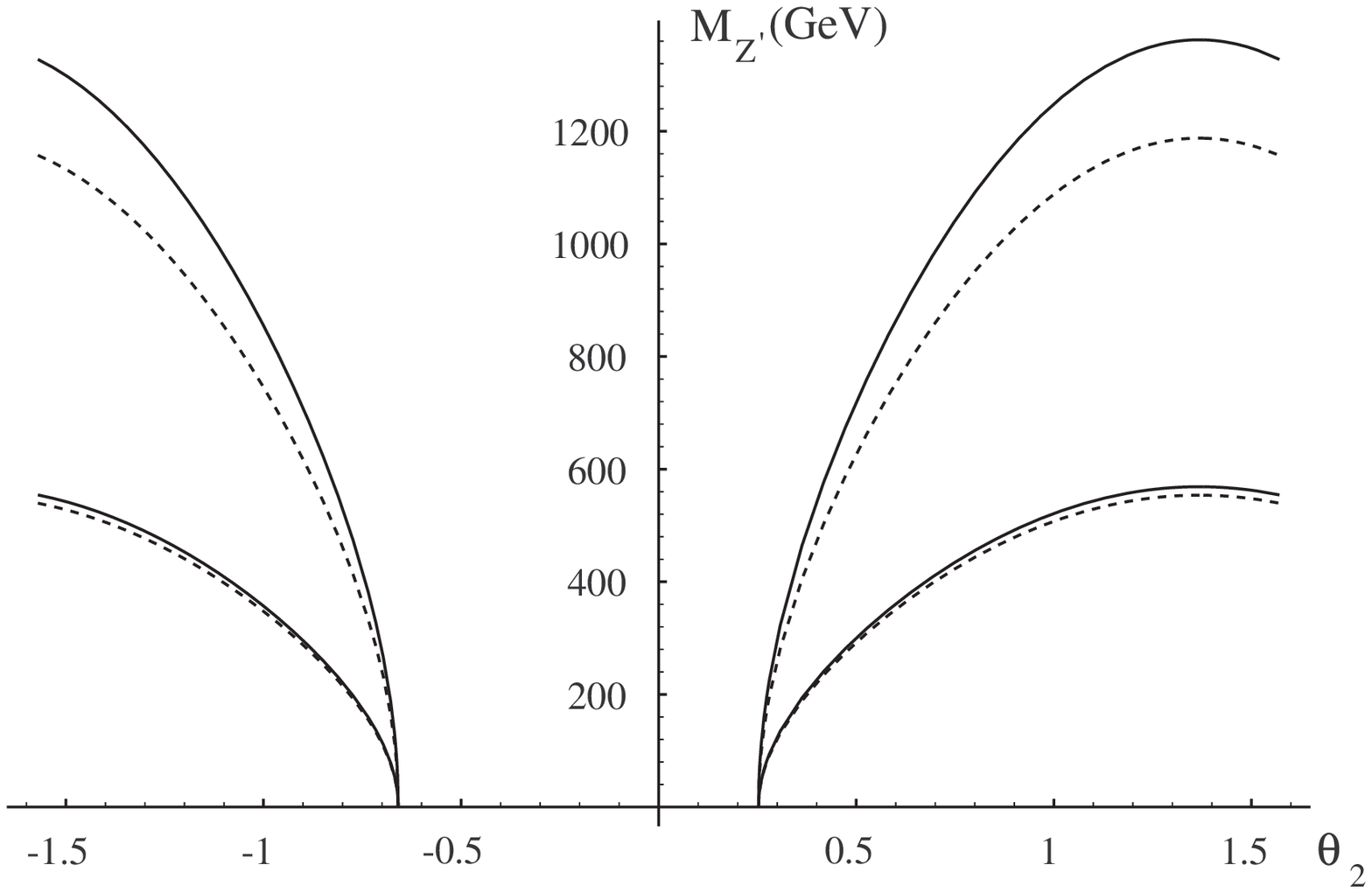}}
\noindent
\caption{
The 95\% CL lower and upper bounds for $M_{Z^\prime}$ for the
extra-U(1) models versus $\theta_2$. The continuous and the
dashed lines correspond to $m_H=70~GeV$ and $m_H=300~GeV$
respectively.}
\label{fig1}
\end{figure}
\end{document}